\documentclass[apj]{emulateapj}
\usepackage{natbib,apjfonts}

\shorttitle{THE DOUBLE CLUSTER ABELL 1758}
\shortauthors{DAVID \& KEMPNER}

\slugcomment{Submitted to the Astrophysical Journal.}

\begin{document}


\title{Chandra and XMM-Newton Observations of the Double Cluster Abell 1758}


\author{Laurence P. David and Joshua Kempner}
\affil{Harvard-Smithsonian Center for Astrophysics, 60 Garden St.,
Cambridge, MA 02138}
\email{david@cfa.harvard.edu}


\begin{abstract}
Abell 1758 was classified as a single rich cluster of galaxies by Abell, but
a ROSAT observation showed that this system consists of two distinct
clusters (A1758N and A1758S)
separated by approximately $8\arcmin$ (a projected separation
of 2~Mpc in the rest frame of the clusters).
Only a few galaxy redshifts have been published for these two clusters,
but the redshift of the Fe lines in the Chandra and XMM-Newton spectra shows that
the recessional velocities of A1758N and A1758S are within 2,100~km~s$^{-1}$.
Thus, these two clusters most likely form a gravitationally bound system,
but our imaging and spectroscopic analyses of the X-ray data
do not reveal any sign of interaction between the two clusters.
The Chandra and XMM-Newton observations show that A1758N and A1758S are both undergoing
major mergers.

A1758N is in the late stages of a large impact parameter merger
between two 7~keV clusters.  The two remnant cores have
a projected separation of 800~kpc.  Based on the measured pressure jumps preceding
the two cores, they are receding from one another at
less than 1,600~km~s$^{-1}$.  The two cores are surrounded by hotter
gas ($\mathrm{kT}=9$--12~keV) that was probably shock heated during the early stages
of the merger.  The gas entropy
in the two remnant cores is comparable with the central entropy observed in
dynamically relaxed clusters, indicating that the merger-induced shocks stalled
as they tried to penetrate the high pressure cores of the two merging systems.
Each core also has a wake of low entropy gas indicating that this gas
was ram pressure stripped without being strongly shocked.

A1758S is undergoing a more symmetric (lower impact parameter) merger between
two 5 keV clusters.  The two remnant cores are nearly coincident as
seen in projection on the sky.  The two cores are surrounded by hotter
gas (9--11 keV) which was probably shock heated during the merger.  Based on
the pressure jumps preceding the two cores, they must have a relative velocity
of less than 1,400~km~s$^{-1}$. Unlike A1758N, there is no evidence for
wakes of low entropy gas.
\end{abstract}





\section{Introduction}

Chandra observations over the past few years have shown that clusters
of galaxies are dynamically complex systems.  These observations
have shown that the hot gas in clusters is frequently perturbed by
cluster mergers and nuclear outbursts and motion of the central
dominant galaxy. Cluster mergers
generate hydrodynamic shocks, cold fronts (contact discontinuities),
and filaments (Markevitch et al.\ 2000;
Vikhlinin, Markevitch, \& Murray 2001a; Markevitch et al.\ 2002; Mazzotta et al.\ 2002;
Mazzotta, Fusco-Femiano, \& Vikhlinin 2002; Kempner, Sarazin, \& Ricker 2002).
Residual motion or ``sloshing'' of the central dominant galaxy in clusters can
also produce cold fronts and possibly wakes of stripped or cooled gas
(Fabian et al.\ 2001; Markevitch, Vikhlinin, \& Mazzotta 2001).
These discoveries have helped illuminate the roles of
magnetohydrodynamics and transport processes during cluster mergers.
The observed sharpness of cold fronts shows that thermal conduction is highly
suppressed in these regions (Ettori \& Fabian 2000; Vikhlinin,
Markevitch, \& Murray 2001a).
The robustness of merging cores demonstrates that the growth of
hydrodynamic instabilities is also suppressed, possibly by the compression
of ambient magnetic fields along cold fronts and the subsequent increase in
magnetic surface tension (Vikhlinin, Markevitch, \& Murray 2001b).
A full understanding of recent
Chandra observations of clusters provides a significant challenge
for future numerical simulations, which must resolve structure
on scales of a few kpc and include the effects of magnetic fields.

While Abell 1758 was classified as a single cluster by Abell (1958), Rosat images
show that there are two distinct clusters (A1758N and A1758S) separated
by approximately $8\arcmin$ (Rizza et al.\ 1998).  Abell 1758N is a hot
($\rm{kT} \sim$~9--10~keV; Mushotzky \& Scharf 1997) and X-ray luminous cluster
($L_{bol}=2.9 \times 10^{45}$~ergs~s$^{-1}$; David, Jones, \& Forman 1999).
No temperature information is available for A1758S in the literature, but
it is only slightly less luminous than A1758N
($L_{bol}=2.0 \times 10^{45}$~ergs~s$^{-1}$; David, Jones, \& Forman 1999).
The Rosat HRI image shows that both A1758N and A1758S are highly disturbed systems
(Rizza et al.\ 1998).  A1758N has two X-ray peaks separated by approximately $3\arcmin$,
while the centroid of A1758S is offset from the centroid of the large scale emission by
approximately $1\arcmin$.  A1758N hosts one of the most
powerful NAT radio sources known (O'Dea \& Owen 1985, Rizza et al.\ 1998).
Head-tail and NAT radio galaxies are most common
in clusters with perturbed X-ray morphologies and are probably produced
as radio galaxies move through cluster atmospheres at high velocities
(Burns et al.\ 1994).   One of the X-ray peaks in A1758N is centered on
a diffuse radio halo (Kempner \& Sarazin 2001), which is further evidence for
a recent major merger.

\begin{figure*}
\plotone{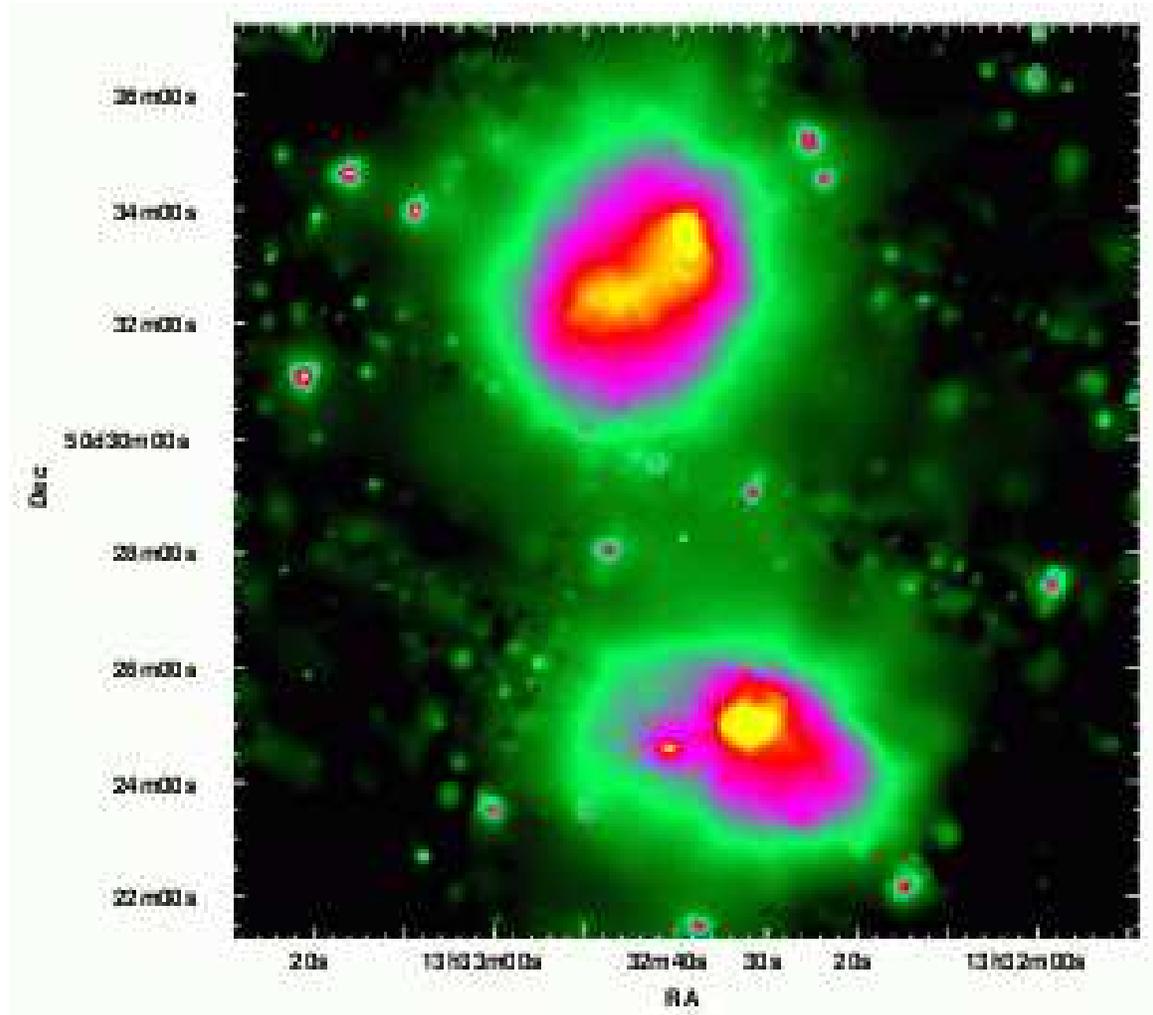}
\caption{Adaptively smoothed, exposure corrected mosaic of the 3 XMM-Newton EPIC
cameras.  The image spans $16\arcmin$ on a side (4.1~Mpc at the redshift of the cluster).
}\label{fig:xmm_mosaic}
\end{figure*}

In this paper, we present Chandra and XMM-Newton observations of A1758.
By utilizing Chandra's high spatial resolution and XMM-Newton's
large throughput, we can search for signs of interaction between
A1758N and A1758S and investigate the present status of the mergers
in the two systems.
This paper is organized in the following manner.  In $\S 2$ we discuss
the details of our Chandra and XMM-Newton data analysis.   Section 3 contains
a discussion about the large scale X-ray properties of A1758 and a search
for any interaction between A1758N and A1758S.  We discuss the dynamic states of
A1758N and A1758S separately in sections 4 and 5.  Our proposed merger
scenarios for the two clusters are summarized in $\S 6$.
We assume $\mathrm{H_0}=70$ km s$^{-1}$ Mpc$^{-1}$, $\Omega_{M}=0.3$,
and $\Omega_{\Lambda}=0.7$ throughout the paper.
At a redshift of 0.279, the luminosity distance to A1758 is
1,430~Mpc, and $1\arcsec = 4.24$~kpc.

\section{Data Reduction}

\subsection{Chandra Data}

Abell 1758N was observed with Chandra on August 28, 2001 for 
58,314s with ACIS-S in VF mode.  The cluster was positioned on the center of the 
back illuminated S3 chip to provide better spatial coverage.  We 
applied the same screening criteria on the level 1 data products
as those used to generate the publically available blank field background files~$^1$,
including the additional screening for charged particles that is possible
in VF mode.  All calibration products used in our analyses are from CALDB 2.23.
Since A1758N essentially covers
the entire S3 chip, we used the data acquired on the back illuminated S1 
chip to remove periods when the background rate
exceeded the average rate by more than 20\%, which left 42,521s of data.
However, even after the short period charged particle flares are removed, the 
average 0.3--10~keV background rate in the S1 chip
is still 18\% higher than the nominal rate and is constant during the
entire observation. The count rates in the front illuminated 
I2 and I3 chips are consistent with the nominal background rates.
This observation (OBSID 2213) was mentioned in Markevitch et al.\ (2003)
as an example of a long period soft flare of charged particles, i.e., a flare that
is only detected in the back illuminated chips.  Markevitch et al.\ found
that such flares can be modeled as a power-law with an index
of 0.15 and an exponential cut-off at 5.6 keV in PHA space. Due to 
the exponential cut-off in the spectra of soft flares, they do not
affect the count rate at high energies which can still be used
to normalize the quiescent background rate.
The count rate between pha channels 2500 and 3000 in
the screened S3 data is 9.5\%
higher than that in the background file.  We therefore
adjusted the exposure time in the background file to give consistent rates
at high energies.

To determine the normalization of the soft flare component, we 
extracted spectra from the corner of the S3 chip 
and the same region on the blank sky observation. The difference 
between these two spectra was then fit with a cut-off power-law 
model in PHA space (i.e., CUTOFFPL/b in XSPEC).
The power-law index and cut-off energy in the CUTOFFPL model were
fixed to the values found by Markevitch et al.\ Only the normalization
was allowed to vary.  All spectral analysis
of ACIS-S3 data presented below includes a CUTOFFPL/b model with the normalization
determined from the area of the source extraction region.  Separate photon-weighted
response and effective area files are generated for each spectrum. The 
effective area files are corrected for the effects of contamination on 
ACIS using the corrarf task.
All ACIS images presented in this paper only include photons with energies 
between 0.3--6.0~keV. By excluding higher energy photons, we eliminate 
approximately 50\% of the S3 background at the expense of only 3\% of 
the emission from a 9~keV thermal source.  In addition to the 
nominal background rate, the soft flare is also removed from all 
imaging analysis using the best fit normalization of the CUTOFFPL model 
derived above. 

\bigskip
\bigskip
\subsection{XMM-Newton Data}

Abell 1758 was observed by XMM-Newton on Nov. 11--12, 2002.  All XMM-Newton 
analysis was done 
with SAS v5.4.1.  The data from all three EPIC cameras were first reprocessed using 
the tasks emchain and epchain.  The data were then screened 
using the standard patterns and flags specified in ``An Introduction
to XMM-Newton Data Analysis''.  To screen out periods of enhanced background,
we generated light curves for each
detector in the 10--15 keV band.
We then performed a successive $2 \sigma$ clipping of the data which yielded
49435s, 49185s, and 43471s of screened data for 
the MOS1, MOS2, and PN cameras, respectively. We only included events with
FLAG=0 in our spectral analysis.  The large field of view of the EPIC cameras
allowed us to extract background spectra from the same data sets.  The background
spectrum from all 3 cameras was extracted from within a $2^{\prime}$ radius circle
located $11^{\prime}$ due north of A1758N.
Separate response and effective area files were generated for each EPIC spectra using
the SAS tasks rmfgen and arfgen.  A detector map was used to weight each
effective area file.

\footnote{http://cxc.harvard.edu/contrib/maxim/bg/index.html}

\begin{figure}
\epsscale{1.11}
\plotone{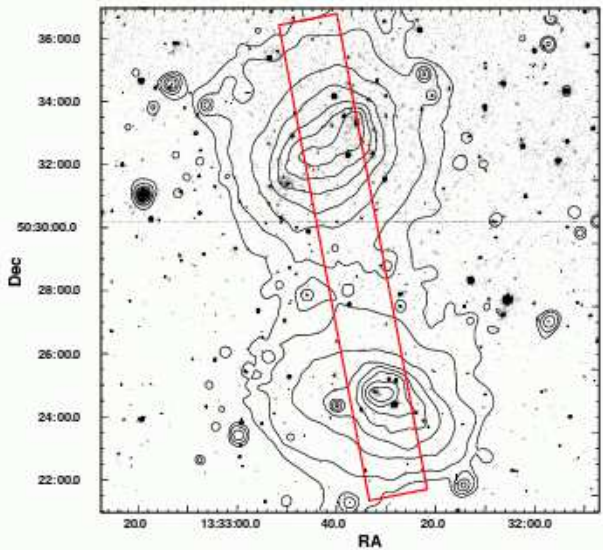}
\caption{Contours of the adaptively smoothed XMM-Newton mosaic shown in 
Fig.~\ref{fig:xmm_mosaic} overlayed on the DSS image.}\label{fig:dss_xmm_mosaic}
\end{figure}

\begin{figure}
\plotone{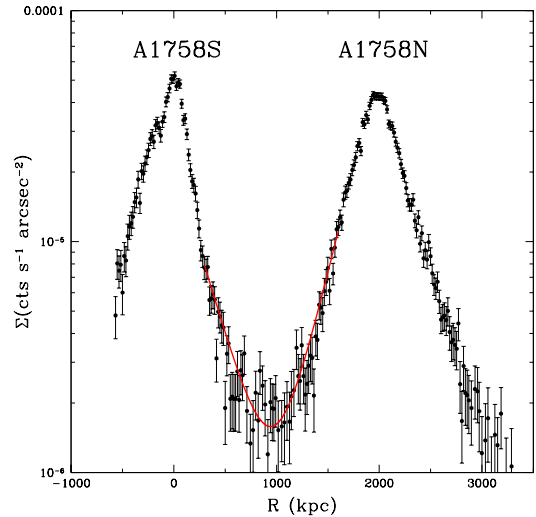}
\caption{The projected, background-subtracted, exposure-corrected, 0.3--7.0 keV
surface brightness profile of the PN data within the rectangular region
shown in Fig.~\ref{fig:dss_xmm_mosaic}. The data were binned into a minimum
of 25 net counts per data point. The red curve is the sum of the best-fit power-law
surface brightness profiles to each cluster.}\label{fig:inter}
\end{figure}

\section{Large Scale Properties of the Double Cluster A1758}

An adaptively smoothed, exposure corrected mosaic of the 3 XMM-Newton 
EPIC images is shown in Fig.~\ref{fig:xmm_mosaic}.  A1758N and A1758S 
are clearly visible in this figure with a projected separation of 
approximately 2.0 Mpc at the redshifts of the clusters.  For comparison, the virial 
radii of the two clusters estimated below are 2.6 Mpc and 2.2 Mpc for 
A1758N and A1758S, respectively. Fig.~\ref{fig:xmm_mosaic} clearly shows
the presence of X-ray emission between the two clusters. 
This emission is analyzed in detail below to search for any signs of interaction
between the two systems.  Both A1758N and A1758S contain significant substructure.
A1758N contains 
two subclusters separated along a line running from SE to NW.  A1758S has 
a large subcluster off-set towards the NE from the main component of the cluster.  
The source to the east of A1758S has a spatial extent consistent with that of a 
point source.  Contours of the adaptively smoothed XMM-Newton mosaic
are overlayed on the DSS image in Fig.~\ref{fig:dss_xmm_mosaic}.
This figure shows that the large scale contours of A1758S are roughly 
centered on the brightest galaxy in the cluster, while the subcluster toward 
the NE in A1758S is centered on a group of fainter galaxies.  
The subcluster in A1758N off-set toward the NW is located near the center of 
a line of bright galaxies, while the SE subcluster in A1758N is 
off-set toward the NW from a fainter group of galaxies. The X-ray morphologies
of A1758N and A1758S are discussed in more detail below.

\begin{deluxetable*}{lcccccc}
\tabletypesize{\scriptsize}
\tablecaption{Global X-Ray Properties of A1758N and A1758S}
\tablehead{
\colhead{} & \colhead{} & \colhead{kT} & \colhead{Z} & \colhead{F$_{0.5-10.0~ \rm{keV}}$} & \colhead{L$_{0.5-10.0~ \rm{keV}}$} & \colhead{} \\
\colhead{Cluster} & \colhead{Detectors} & \colhead{(keV)} & \colhead{($Z_{\odot}$)} & \colhead{(ergs~cm$^{-2}$~s$^{-1}$)} & \colhead{(ergs~s$^{-1}$)} & \colhead{$\chi^2$/DOF} \\
\vspace{-0.11in}
}
\startdata
A1758N & ACIS-S3 & 9.0 (8.4--9.9) & 0.30 (0.23--0.36) & $7.03 \times 10^{-12}$ & $1.73 \times 10^{45}$  & 307/306 \\
A1758N &  MOS1+MOS2+PN & 8.2 (8.0--8.6) & 0.28 (0.24--0.33) & $6.75 \times 10^{-12}$ & $1.66 \times 10^{45}$ & 1891/1815 \\
A1758S & MOS1+MOS2+PN & 6.4 (6.0--6.7) & 0.22 (0.15--0.28)& $4.49 \times 10^{-12}$ & $1.49 \times 10^{45}$ & 1738/1477 \\
\vspace{-0.08in}
\enddata

\tablecomments{Results of fitting the XMM-Newton and Chandra spectra extracted within 1~Mpc radius 
apertures centered on A1758N and A1758S. All spectra were binned to a minimum 
of 20 counts per bin and fit to an absorbed MEKAL model with the absorption fixed 
to the galactic value of $1.05 \times 10^{20}$~cm$^{-2}$. This table gives the 
best-fit gas temperature (kT), abundance relative to the solar value (Anders \& Grevesse 1989),
unabsorbed 0.5-10.0~keV flux and luminosity, minimum $\chi^2$ and degrees 
of freedom (DOF).  All errors are given at the 90\% confidence limit.}
\end{deluxetable*}

To search for signs of interaction between A1758N and A1758S, we extracted the counts
from the PN image within the rectangular region shown in 
Fig.~\ref{fig:dss_xmm_mosaic}. Only the PN image is used in this analysis
due to its lower background compared to the MOS detectors.
A projected surface brightness profile
was then computed by projecting the background-subtracted and exposure-corrected counts
between 0.3 and 7.0 keV onto the longer side of the rectangle 
(see Fig.~\ref{fig:inter}). The emission from all point sources between the 
two clusters was removed from the analysis. If the two clusters are interacting,
then the gas at the outskirts of the clusters should be compressed 
and produce an excess of emission above the projected emission from the two systems.
Since the emission from the clusters is not spherically symmetric, we 
fit the surface brightness in the regions from 300 kpc to 600 kpc in A1758S and 
from 1300 kpc to 1700 kpc in A1758N to simple power-laws, where zero 
is centered on A1758S.  These radii correspond 
to those shown in Fig.~\ref{fig:inter}.  The sum of the two best fit power-laws 
is shown as the red line in Fig.~\ref{fig:inter}.
Our analysis indicates that 
there is no statistically significant excess emission between the two clusters 
above that expected from a simple projection of the two systems.  
We also generated a hardness ratio profile using the PN data 
to search for signs of compressional heating, but
the resulting hardness ratio does not vary significantly 
between the two clusters.

\begin{figure}
\plotone{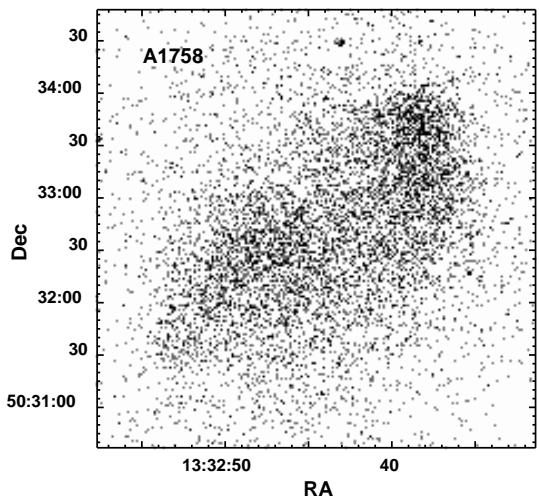}
\caption{Raw 0.3--6.0~keV ACIS-S3 image of A1758. This figure spans 1.08~Mpc on a
side in the rest frame of the cluster.}\label{fig:raw}
\end{figure}

The global X-ray properties of the two clusters were determined by 
extracting spectra within 1 Mpc radius apertures.
We extracted spectra from all 3 EPIC cameras for both clusters and
the ACIS-S3 spectra for A1758N.  This is the largest extraction
region that can be used without any overlap between the two clusters
and also fit within the ACIS-S3 field-of-view. These spectra
were then fit to an absorbed single temperature MEKAL model.
The three EPIC spectra were fit simultaneously in the 0.5--10.0~keV 
energy band and the ACIS-S3 spectrum was fit separately in the 0.7--10.0~keV energy band.
Table 1 shows that there
is good agreement between the XMM-Newton and Chandra spectral analyses of A1758N. Our 
temperature measurements for A1758N also are in good agreement with 
a previous analysis of ASCA data by Mushotzky \& Scharf (1997)
who found $kT= 8.5$--$12.5$~keV (90\% errors).  Even though these clusters
are undergoing major mergers at the present time, the scaling between their
temperatures and luminosities is consistent with the self-similar 
solution, $L \propto T^2$. Simulations of cluster merging by Randall, Sarazin, \& Ricker (2002)
and Rowley, Thomas, \& Kay (2003) find that even though the temperature and
luminosity of a cluster varies significant during a merger, it still
follows a $L \propto T^2$ relation.
Using the cluster scaling relations in Bryan \& Norman (1998), 
our temperature estimates, and our adopted cosmology (see the end of $\S 1$), 
we estimate virial masses of $1.0 \times 10^{15}$~M$_{\odot}$ and 
$1.6 \times 10^{15}$~M$_{\odot}$ for A1758S and A1758N, respectively.
These calculations are only used as rough guides since the
on-going mergers likely produce emission-weighted temperatures
that differ from the virial temperatures of the clusters,
(see the simulations of Randall, Sarazin, \& Ricker 2002), but the masses
are probably accurate to within a factor of two.

Since there are so few published galaxy redshifts for A1758N and A1758S,
we allowed the redshift to vary in our spectral analysis of the EPIC data.
We obtained best-fit redshifts of z=0.2756 (0.2742--0.2806) for A1758N 
and z=0.2750 (0.2666--0.2808) for A1758S ($1 \sigma$ errors). 
This gives a velocity difference less than 
2,100 km~s$^{-1}$ between the two clusters at $1 \sigma$ confidence. 
The latest calibration summary document released by the XMM-Newton
team (available at http://xmm.vilspa.esa.es) states that the gain uncertainties 
in the EPIC cameras using SAS 5.4.1 are approximately 10 eV at all 
energies.  At the Fe-K line, this corresponds to a
velocity uncertainty of only $450$~km~s$^{-1}$.  Using the above estimates on the 
virial mass and radius of A1758N, the infall velocity at the virial radius 
of 2.6 Mpc in A1758N 
is 2,300~km~s$^{-1}$, which is consistent with the observed velocity difference.
Thus, even through there are no direct signs of interaction in the X-ray 
data between the two clusters, their proximity to one another and small
velocity difference are consistent with being a gravitationally bound system.

\begin{figure}
\epsscale{1.09}
\plotone{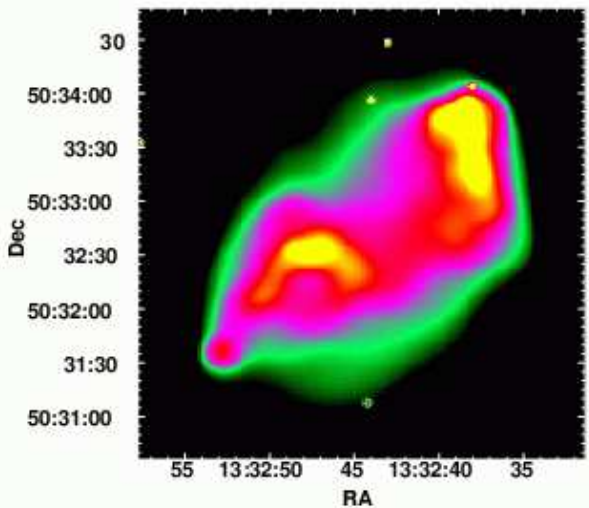}
\caption{Adaptively smoothed, background-subtracted, and exposure-corrected 
ACIS-S3 image of A1758N.}
\label{fig:adapt}
\end{figure}

\section{Dynamic State of the A1758N Cluster}

\noindent
A1758N has a very complex X-ray morphology with several sharp features surrounding 
two merging subclusters (see the raw Chandra image in Fig.~\ref{fig:raw}). 
The ACIS-S3 image shows that A1758N has a sharp front along the western edge of the
cluster and two angular shaped edges bounding the
NW and SE subclusters. There is also a depression in the surface brightness
between the two subclusters.  The adaptively smoothed ACIS-S3 image in 
Fig.~\ref{fig:adapt} shows the cores of the two merging 
systems more clearly with a separation of 
approximately 800~kpc.  With Chandra's higher spatial resolution, we can identify 
the center of the SE subcluster with the group of galaxies visible 
in Fig.~\ref{fig:dss_xmm_mosaic}.

\begin{figure}
\plotone{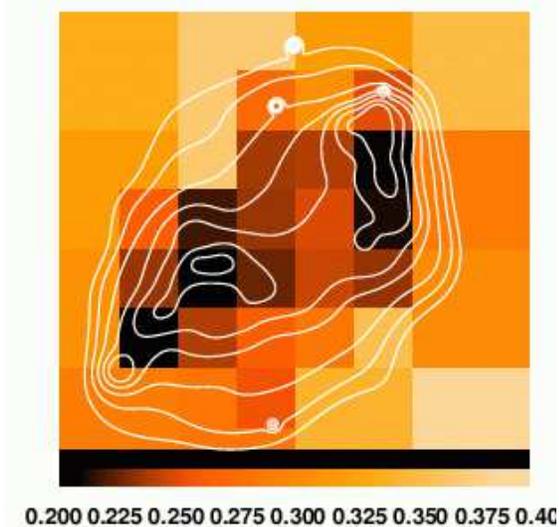}
\caption{X-ray contours of the adaptively smoothed ACIS-S3 image of A1758N overlayed
on an adaptively binned hardness ratio map with a maximum relative error of 10\% 
using a soft band of 0.3--2.0~keV and a hard band of 2.0--6.0~keV. 
The image spans $4^{\prime}$ on a side.
The tic marks on the color bar are spaced at approximately $1 \sigma$ intervals.}\label{fig:HR}
\end{figure}

We generated an adaptively binned hardness ratio map using the technique in 
Sanders \& Fabian (2001) with energy bands of 0.3--2.0~keV and 2.0--6.0~keV.  An 
energy cut of 2.0~keV gives a reasonable compromise between temperature sensitivity 
(which requires a high energy cut) and photon statistics (which requires
a low energy cut).  Fig.~\ref{fig:HR} displays the X-ray contours from the 
adaptively smoothed ACIS-S3 image overlayed on the
adaptively binned hardness ratio map. The data is binned so that the maximum relative 
errors in the hardness ratios are 10
brightest regions have the softest emission, and hence must contain the 
lowest entropy gas.  The hardness ratios of the gas in the darkest regions
adjacent to the two subclusters are approximately $3 \sigma$ less than the 
hardness ratio of the gas in the remainder of the cluster core.
As a subcluster moves through another cluster, the pressure 
and density gradients around the subcluster should be steepest in the direction of 
motion. Examining the X-ray contours of A1758N suggests that the NW system is presently 
moving toward the north while the SE subcluster is presently moving toward the SE. 
The diffuse radio halo in A1758N (Kempner \& Sarazin 2001) is centered
on the NW subcluster and is also elongated in a north-south direction.
The low entropy gas trailing these two systems is thus most likely gas that
has been stripped into the wakes of the two merging subclusters. 

\begin{figure}
\epsscale{1.1}
\plotone{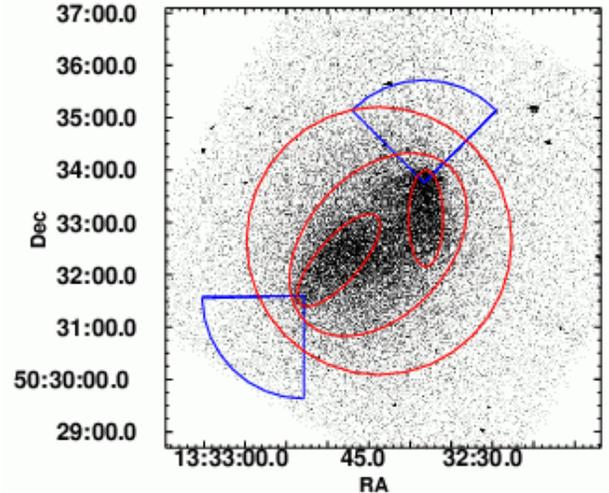}
\caption{The regions in A1758N used for spectral analysis (red) and 
generating surface brightness profiles (blue) across the leading edges of the 
NW and SE subclusters.} \label{fig:regions}
\end{figure}

Using the X-ray morphology and hardness ratio map of A1758N as guides, 
we extracted ACIS-S3 and EPIC spectra from the 4 regions shown in red in
Fig.~\ref{fig:regions}. We then fit the spectra to absorbed single temperature
MEKAL models and the results of our spectral analysis are given in Table 2.  The ``halo'' 
referred to in Table 2 corresponds to the region beyond the largest ellipse but 
within the large circle in Fig.~\ref{fig:regions}. 
The ``core'' covers the region within the largest ellipse but excludes emission from 
the NW and SE wakes (the two smaller ellipses). Based on the spectral analysis 
presented in Table 2, the halo gas is significantly hotter than the gas 
in the wakes.  A1758N is undoubtedly undergoing a major merger and the high
gas temperature in the halo suggests that any shocks generated during 
the merger have already propagated throughout most of the cluster.  

\begin{deluxetable*}{lccc|ccc}
\tabletypesize{\scriptsize}
\tablecaption{Spectral Analysis Results for A1758N}
\tablehead{
\colhead{} & \multicolumn{3}{c|}{ACIS-S3} & \multicolumn{3}{c}{MOS1+MOS2+PN} \\
\colhead{} & \colhead{kT} & \colhead{Z} & \multicolumn{1}{c|}{~} & \colhead{kT} & \colhead{Z} & \colhead{} \\
\colhead{Region} & \colhead{(keV)} & \colhead{($Z_{\odot}$)} & \multicolumn{1}{c|}{$\chi^2$/DOF} & \colhead{(keV)} & \colhead{($Z_{\odot}$)} & \colhead{$\chi^2$/DOF} \\
\vspace{-0.11in}
}
\startdata
A1758N (NW Wake) & 7.0 (5.9--8.9) & 0.32 (0.10--0.57) & 119/122  & 7.2 (6.7--7.6) & 0.35 ( 0.25--0.45) & 651/592 \\
A1758N (SE Wake) & 6.7 (5.8--8.3) & 0.27 (0.02--0.54) & 73/85 & 6.6 (6.0--7.1) & 0.45 (0.31--0.60) & 317/370 \\
A1758N (Core) & 8.7 (8.1--9.4) & 0.36 (0.22--0.50) & 330/336 & 7.2 (6.9--7.4) & 0.26 (0.21--0.30) & 1536/1378 \\
A1758N (Halo) & 10.5 (9.3--12.1) & 0.11 ($<$0.33) & 472/490 & 9.8 (9.1--10.6) & 0.25 (0.16--0.36) & 1411/1387 \\
\vspace{-0.08in}
\enddata
\tablecomments{Spectral fitting results for the four regions in A1758N illustrated in 
Fig.~\ref{fig:regions}. See the notes to Table 1 for further details.}
\end{deluxetable*}

\begin{deluxetable*}{lccc}
\tablecaption{XMM-Newton Spectral Analysis Across the NW Edge in A1758N}
\tablehead{
\colhead{}       & \colhead{kT}    & \colhead{Z}             & \colhead{}  \\
\colhead{Region} & \colhead{(keV)} & \colhead{($Z_{\odot}$)} & \colhead{$\chi^2$/DOF} \\
\vspace{-0.11in}
}
\startdata
NW Spheroid& 7.2 (6.4--8.1)(5.9--8.8) & 0.64 (0.33--1.0)(0.15--1.2) & 87/94 \\
NW Inner Power-Law & 8.0 (7.0--8.8)(6.7--9.4) & 0.11 ($<$0.28)($<$0.38) & 184/147 \\
NW Outer Power-Law& 10.8 (9.5--12.0)(9.0--13.1) & 0.17 (0.02--0.32)($<$0.43) & 351/285 \\
\vspace{-0.08in}
\enddata
\tablecomments{Gas temperatures and abundances within the pie slice centered on the NW 
subcluster in A1758N (see Fig.~\ref{fig:regions}).  The regions refer to the
3 model components fit to the surface brightness profile
(see Fig.~\ref{fig:josh_sb}). Both $1 \sigma$ and 90\% errors are given.
See the notes to Table 1 for further details on the spectral analysis.}
\end{deluxetable*}

\begin{figure}
\plotone{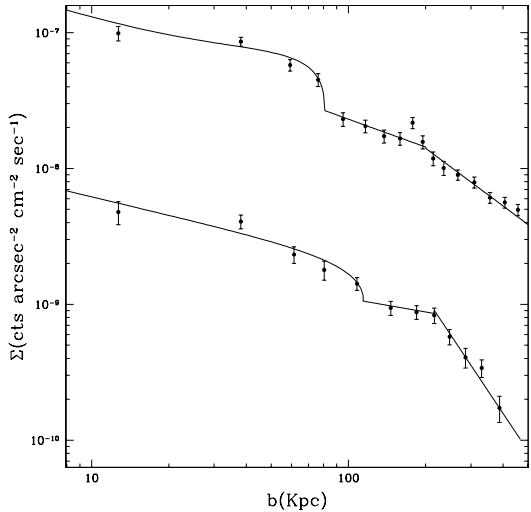}
  \caption{The background-subtracted and exposure-corrected ACIS-S3 0.3--6.0 keV
  surface brightness profiles within the blue pie slices shown in Fig.~\ref{fig:regions}. 
  The NW subcluster is shown in the upper curve and the SE subcluster is shown
  in the lower curve.  The lower curve is off-set by an order 
  of magnitude for clarity.  The best fit truncated spheriod plus broken power-law 
  models are also shown.}
  \label{fig:josh_sb}
\end{figure}

To investigate the dynamic state of the gas in A1758N further, we extracted surface 
brightness profiles from the ACIS-S3 data across the leading edges of the NW and SE 
subclusters (see the blue pie slices in Fig.~\ref{fig:regions}). The resulting 
background-subtracted and exposure-corrected surface brightness profiles are shown in 
Fig.~\ref{fig:josh_sb}.  We first fit the profiles with a truncated spheroid 
given by $S(d) = A \sqrt{R} \sqrt{d}$, where $R$ is the radius of curvature of 
the cold front and $d$ is the distance into the cold gas from the cold front
(eq. A5 in Vikhlinin, Markevitch, \& Murray 2001a), plus a power-law model,
but as can be seen in Fig.~\ref{fig:josh_sb}, the outer regions are not well 
represented by a single power-law.  We then tried a truncated spheroid plus a broken 
power-law model which produced satisfactory fits. The best-fit radii
for the truncated spheroids are 80~kpc and 115~kpc for the NW 
and SE subclusters, respectively.
The best-fit power laws correspond to $\beta$ values of $0.40 \pm 0.03$ and 
$0.60 \pm 0.02$ in the NW system and $0.22 \pm 0.05$ and $0.64 \pm 0.04$ 
($1 \sigma$ errors) in the SE system.  The outer slopes in the two
clusters are typical of hot relaxed clusters, 
while the inner slopes are considerably flatter.  Deprojecting the surface brightness 
profiles give central electron number densities of $0.017 \pm 0.0003$~cm$^{-3}$ and 
$0.012 \pm 0.0005$~cm$^{-3}$ in the NW and SW subclusters, which again, are typical 
densities 
for rich clusters. The density jumps preceding
the NW and SE subclusters (i.e., at the edge of the truncated
spheroidal component) are $\rho_2/\rho_1=1.6 \pm 0.2$ and $1.5 \pm 0.2$
($1 \sigma$ errors).

Based on the success of the 3 component model in fitting the 
surface brightness profiles across the NW and SE edges,
we extracted ACIS-S3 and EPIC spectra within the three regions covered by the 
different components of the model.  The photon statistics of the 
ACIS-S3 data within these regions were not sufficient to
place strong constraints on the gas temperature distribution.
We also found that the spectral analysis of the EPIC data across the 
SE edge provided little information on the temperature distribution,
so we only present the results for the EPIC data across the NW edge in Table 3. 
This table shows that the gas temperatures within 
the truncated spheroid and inner power-law regions are comparable, while the 
gas in the outer power-law region is significantly hotter.  There is 
also some evidence for a higher abundance within the spheroidal component.
The three components in the model surface brightness are suggestive
of an inner core bounded by a cold front or contact discontinuity,
a region of shocked gas bounded by a stand-off bow shock, and pre-shocked
gas.  However, in such a situation, the core and pre-shocked gas should
be the coldest and the shocked gas the hottest, which is contrary to the 
observations.  Since the gas remains hot to large distances in front
of the NE subcluster, the shock must have already propagated throughout
most of the cluster. This will be discussed in more detail below.  

\subsection{Proposed Merger History for A1795N}

Based on past theoretical studies of merging clusters and the
disruption of cooling cores during merging
(e.g., Fabian \& Daines 1991; Pearce, Thomas, \& Couchman 1994;
Ricker \& Sarazin 2001; Gomez et al.\ 2002, Randall, Sarazin, \& Ricker 2002), 
the merging process 
can be broken down into several stages: 1) the initial encounter phase 
during which shocks develop in the cluster outskirts,
2) the shocks stall as they try to penetrate the 
high pressure low entropy cores, generating the largest
entropy jump in the outer gas, which had the highest initial entropy,
3) as the low entropy cores approach one another through shocked gas, 
they are continuously stripped as long as the ram pressure exceeds 
the thermal pressure within the cores, 4) stripping continues until 
all the gas beyond the core radius of the dark matter distribution has been
stripped and then the remainder of the gas is stripped in bulk 
due to the nearly constant binding energy per unit mass of the core gas,
5) for off-axis mergers, the cores can survive beyond the point of closest 
approach, after which ram pressure stripping becomes less efficient and 
the surviving cores expands adiabatically,
6) the chaotic entropy distribution of the merged cluster drives
large scale convection and some of the
the shock energy that was deposited in the cluster outskirts is transported
into the core gas, 7) eventually the two dark matter halos merge, and the lowest 
entropy gas settles into the minimum of the gravitational potential.

The X-ray contours, hardness ratio map, and surface brightness profiles of
A1758N are consistent with a scenario in which the NW subcluster is presently 
moving toward the north and the SE subcluster is presently moving 
toward the SE (i.e., stage 5 above).  
The NW subcluster has been stripped down to its inner 80~kpc and the SE subcluster 
to its central 115~kpc.  The survival of the two cores indicates that the core radii
of the dark matter distributions in the merging systems are less
than about 100 kpc.  This is consistent with estimates deduced
from Chandra observations of relaxed clusters (David et al.\ 2001,  
Arabadjis, Bautz, \& Garmire 2002, Allen, Schmidt, \& Fabian 2002).
Based on the central densities and temperatures derived above,
the central gas entropies are
120 and 110~keV~cm$^2$ for the SE and NW subclusters. These values
are consistent with the central regions of relaxed clusters (e.g., Lloyd-Davies,
Ponman, \& Cannon 2000) and show that this gas was not
strongly shocked during the merger.  
The hardness ratio map shows that the emission from the
gas in the wakes is also soft, indicating that this gas was stripped 
without being strongly shocked. The detached nature of the wake in the SE 
subcluster, as evident in Fig.~\ref{fig:adapt}, indicates that ram pressure 
stripping is lessening, as it should after the point of closest approach.

\subsection{Merger Kinematics for A1758N}

\begin{figure}
\plotone{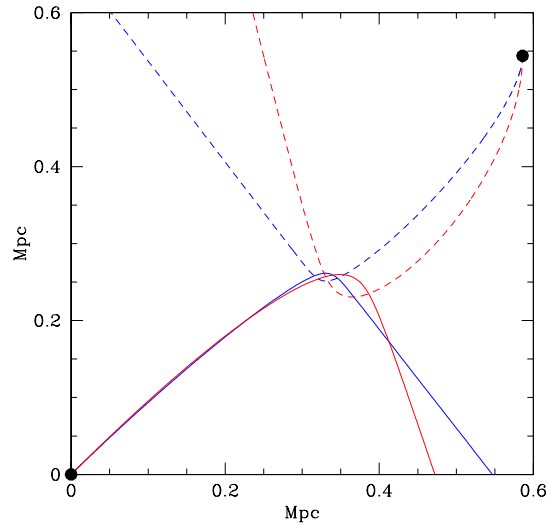}
  \caption{Trajectories of the two subclusters in A1758N assuming present
  velocities of 500~km~s$^{-1}$ (blue) or 1000~km~s$^{-1}$ (red).
  Their present positions are indicated with filled circles. The trajectory of 
  the SE subcluster is shown with a solid line and the trajectory of the NW subcluster is shown
  with a dashed line.}\label{fig:orbit}
\end{figure}

Combining the derived density and temperature jumps across the NW edge
gives a pressure jump of $1.4 \pm 0.3$ ($1 \sigma$ error).  The pressure
in the core should be greater than that in the ambient medium due to 
ram pressure; however, the error on the pressure jump limits the
relative velocity between the two subclusters to be less than 
approximately 1600~km~s$^{-1}$.
The curved shape of the wakes in A1758N shows that the collision of the two systems 
was not head-on.  The SE subcluster appears to have been deflected from a NE
trajectory towards the SE, while the NE cluster appears to have been deflected 
from a western trajectory towards the north. While we cannot
determine the detailed merger kinematics without a high resolution
temperature map, which would require significantly better photon statistics,
we can check the validity of our proposed scheme by
simply integrating the equations of motion backward for the two systems from
their present positions.  To do this we assume: 
1) the SE subcluster is presently moving toward the SE
and the NW subcluster is presently moving north, 
2) the two subclusters can be represented by point masses 
with masses equal to the virial mass of a 7~keV cluster (using the 
scaling relations in Bryan \& Norman 1998), and 3) the 
orbits are in the plane of the sky.
Figure ~\ref{fig:orbit} shows the resulting trajectories assuming 
present velocities for the NW and SE subclusters of 500 or 1000~km~s$^{-1}$
in the plane of the sky.
While this is a highly idealized calculation, both sets of trajectories 
have similar curvatures as the wakes in Fig.~\ref{fig:adapt}. 
Of course, the extended mass distribution in the clusters would soften the deflection 
angle near pericenter.  Fig. ~\ref{fig:adapt}
shows that the wake of the SE subcluster is brightest where the curvature is 
the greatest which occurs at pericenter where ram pressure stripping is
the strongest. 

Several simulations of merging clusters have recently been presented
in the literature that have similar characteristics with A1758N.
Figure 3 in Bialek, Evrard, \& Mohr (2002) illustrates
the trajectory of two subclusters that undergo a deflection
of about $\sim 90^{\circ}$ during their encounter.
The smaller cluster in their simulation has a steep X-ray surface 
brightness profile at its leading edge (i.e., a cold front), and 
is trailed by denser cooler gas that was displaced by ram pressure.
A series of off-axis cluster mergers with different mass ratios
was recently published by Ricker \& Sarazin (2001). In their
simulations with impact parameters of 2$r_s$ and 5$r_s$ (where $r_s$ 
is the scale radius in the NFW dark matter density profile), the clusters 
experience a significant deflection during their encounter.
Their Figure 4 shows that both subclusters are preceded with
shocks as they begin separating.
The present state of the gas in A1795N is probably best reproduced
by the simulations presented in Onura, Kay, \& Thomas (2002).
Clusters 2 and 8 in their Figure 1 are bimodal with the two 
subclusters, identified by clumps of low entropy gas, presently
separating and preceded by cold fronts and shock heated gas.

\subsection{Evolution of the Merger Shocks in A1758N}

There is no evidence for a shock front within 400~kpc 
of either subcluster in A1758N.  The average gas temperature within this 
region has a significantly higher temperature than the gas
in the two merging cores, indicating that any merger shocks induced
during the encounter have already passed through most of the cluster.
As demonstrated by Vikhlinin, Markevitch, \& Murray (2001a),
the strength of a shock can be derived from the ratio of 
the stand-off distance between the cold front and the shock,
$d$, to the radius of curvature of the cold front, $r_{c}$.
For weak shocks, this relation can be approximated by
$d/r_c =0.75({\cal M}^2-1)^{-1}$.
For the NW and SE subclusters, $d/r_c$ is greater than 4.0 and 2.5,
respectively.
These large stand-off distances imply shock strengths ${\cal M} \leq 1.15$
in both systems.  This corresponds to a relative velocity between the 
two systems of less than 1,500~km~s$^{-1}$, which is consistent with 
the observed thermal pressure jump across the cold fronts.
Correcting for projection effects can only increase 
the true stand-off distances to the shocks and hence lower the upper limits
on the present shock strengths.  
The expression for $d/r_c$ used above is only valid in steady-state.
Once the stripped remains of the low entropy cores
pass the point of closest approach, they will decelerate due to 
gravity and drag. In the steady-state solution, the shock stand-off 
distance approaches infinity as the shock decelerates to the sound speed. 
Obviously, since the stripped cores become subsonic in a finite time, 
the shock front must begin trailing behind that predicted by the steady-state solution.
Correcting for this effect would further lower the limits on the shock strengths.

\begin{figure}
\epsscale{1.167}
\plotone{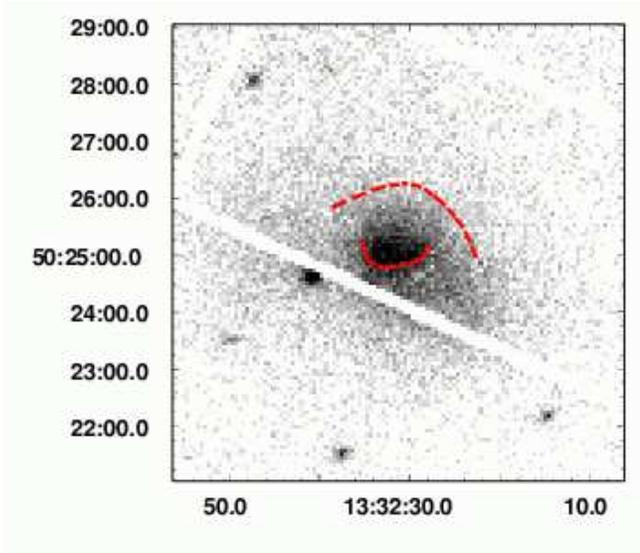}
  \caption{Raw 0.3--7.0~keV PN image of A1758S. This figure spans 2.0~Mpc
  on a side in the rest frame of the cluster. The red dashed curves outline
  the two edges noted in the text.}\label{fig:a1758S_raw}
\end{figure}
 
We have shown that the merger in A1758N is in the late stages of an
off-axis encounter between two massive ($kT \sim 7$~keV) systems.
Any merger shocks induced during this encounter have already
propagated throughout most of the merged system.
Randall, Sarazin, \& Ricker (2002) showed that the final gas temperature after 
the merger of two equal mass clusters is approximately 1.4 times their initial 
temperature.  In A1758N, this would predict a final temperature of about 10 keV
which agrees with the temperature of the gas beyond the two cores. 
Fig.~\ref{fig:josh_sb} shows that the surface brightness of the 
gas preceding the two cold fronts has a broken power-law 
profile, but the spectral analysis of the NW subcluster showed that 
the break was inconsistent with a shock front.  During 
an off-axis encounter, the shock strength, which is determined by the
ratio of post-shocked to pre-shocked thermal pressure, will vary greatly
as the two systems propagate through gas with a broad range of ambient
pressures.  Spectral analysis of the gas preceding the NW subcluster
indicates that the gas closer to the cold front is 
cooler than the gas farther from the cold front.  Such a situation
could arise if the gas that is presently closest to the cold front 
had a higher initial gas pressure, and hence was shocked more weakly,
than the gas presently at larger radii.  If the two clusters
are indeed moving away from one another at the present time,
they should be moving into lower pressure gas which would produce the 
observed temperature distribution of the gas in front of the 
NW subcluster.

\section{Dynamic State of the A1758S Cluster}

\begin{figure}
\plotone{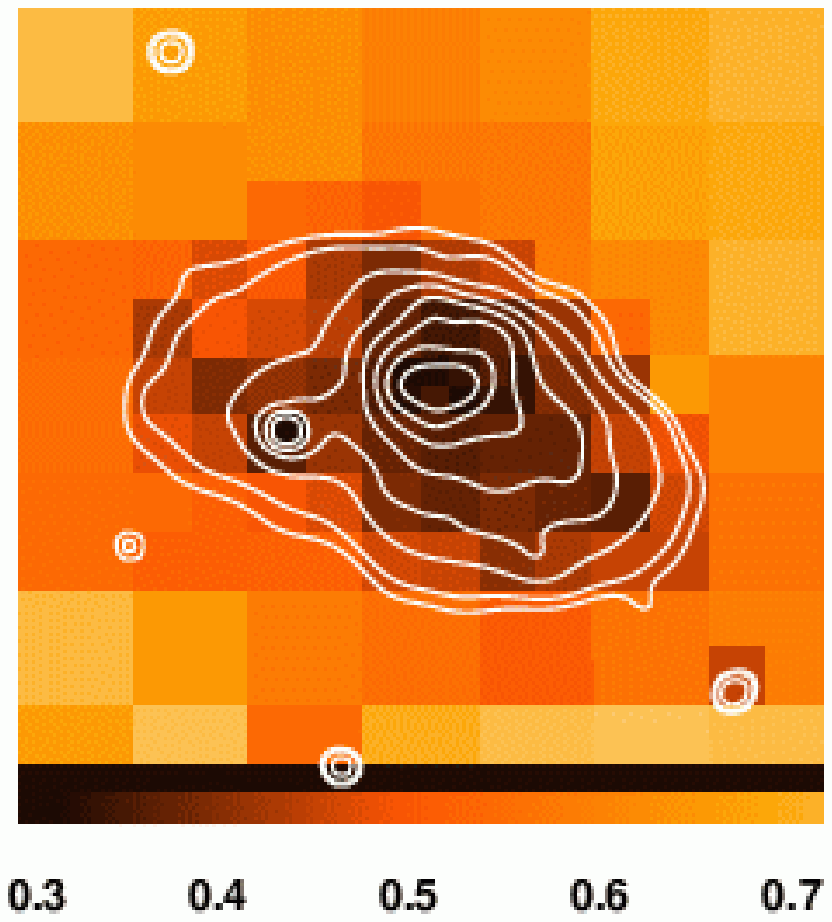}
  \caption{X-ray contours of the adaptively smoothed XMM-Newton Mosaic
  in Fig.~\ref{fig:xmm_mosaic} overlayed on an adaptively binned hardness ratio 
  map with a maximum relative error of 10\% using a soft band of 0.3--2.0~keV and a 
  hard band of 2.0--7.0~keV.  The image spans $7.5^{\prime}$ on a side. 
  The tic marks on the color bar are spaced at 
  approximately $2 \sigma$ intervals.}\label{fig:HR_S}
\end{figure}

The southern cluster in A1758 is also far from dynamical 
equilibrium as can be seen in the raw PN image (see Fig.~\ref{fig:a1758S_raw}).  
With its lower background, the PN data best illustrates the sharp features
in A1758S.  There are two curved edges visible in Fig.~\ref{fig:a1758S_raw}.
The smaller edge toward the south is concave up and the larger edge toward the north
is concave down.
These edges have a similar morphology as those observed in A2142
(Markevitch et al.\ 2000). The orientation of the edges in A1758S
suggest that the merger is occurring along an axis running from NNW to SSE
and has a smaller impact parameter than the merger in A1758N.
An adaptively binned hardness ratio map of A1758S is shown
in Fig.~\ref{fig:HR_S} along with contours derived from the EPIC mosaic 
(Fig.~\ref{fig:xmm_mosaic}).  As in A1758N, the gas within the two edges is
not only the brightest, it is also the coolest, and hence
contains the lowest entropy gas.  Thus, the two edges in A1758S are most
likely cold fronts separating the stripped low entropy cores
of two merging subclusters from higher entropy shock heated gas.

\begin{figure}
\epsscale{1.15}
\plotone{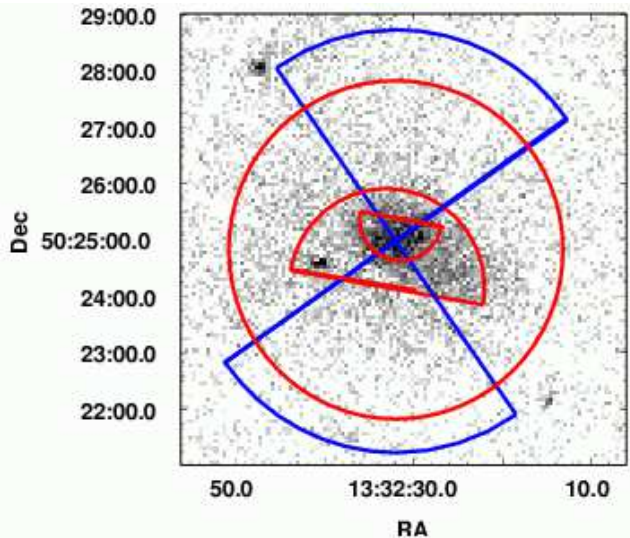}
  \caption{Regions used for spectral and spatial analysis in A1758S.  The two red semi-circles 
  correspond to the small and large cores noted in Table 4 and the region inside
  the red circle but outside the largest semi-circle is noted as the halo.
  The two blue pie slices show the regions used to extract the surface brightness 
  profiles shown in Fig.~\ref{fig:a1758S_josh_sb}.
  The image is the sum of the 3 EPIC cameras in the 0.3--7.0 keV band pass
  and spans 2.0 Mpc on a side.}\label{fig:a1758S_regions}
\end{figure}

As we did for A1758N, we used the observed morphology and hardness ratio
map of A1758S as guides and extracted spectra from all 3 EPIC cameras 
within the regions shown in Fig.~\ref{fig:a1758S_regions} after
excising the emission from all point sources. The results of our spectral 
analysis are presented in Table 4. The small and large cores in Table 4 refer 
to the two semi-circles in Fig.~\ref{fig:a1758S_regions}.  The emission 
from the small core is excised from the emission within the large core.
The halo in Table 4 refers to the region within the circle but outside
the largest semi-circle.
As suggested by the hardness ratio map, the halo gas
is approximately twice as hot as the gas in the two cores.
We also obtained a deprojected gas temperature for the gas within the
small core by subtracting the emission from the surrounding larger core and
obtained essentially the same best fit temperature.

\begin{deluxetable}{lccc}
\tablecaption{XMM-Newton Spectral Analysis for A1758S}
\tablehead{
\colhead{}       &  \colhead{kT}   & \colhead{Z}             & \colhead{} \\
\colhead{Region} & \colhead{(keV)} & \colhead{($Z_{\odot}$)} & \colhead{$\chi^2$/DOF} \\
\vspace{-0.11in}
}
\startdata
Small Core  & 4.6 (4.3-4.9) & 0.26 (0.17-0.35) & 574/414 \\
Large Core  & 5.6 (5.1-6.0) & 0.06 ($<$ 0.14) & 396/418 \\
Halo  & 10.0 (8.6-11.5) & 0.28 (0.09-0.47) & 1019/923 \\
\vspace{-0.08in}
\enddata
\tablecomments{EPIC spectral fitting results for the three regions in A1758S illustrated in 
Fig.~\ref{fig:a1758S_regions}. See the notes to Table 1 for further details of the 
spectral analysis.}
\end{deluxetable}

\begin{figure}
\plotone{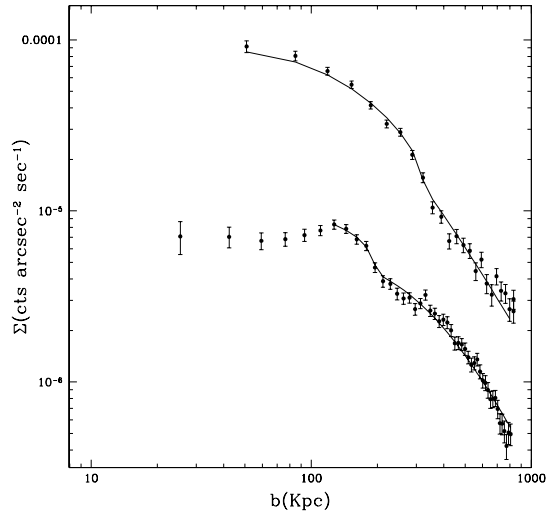}
  \caption{The 0.3-7.0 keV, background-subtracted and exposure-corrected surface brightness profiles
  across the northern (top) and southern (lower) edges in A1758S. Due to the location
  of the chip gaps, different detectors are used to extract the surface brightness
  profile across the two edges.  The profile
  across the northern edge is derived from from the PN data while the profile across the
  southern edge is derived from the sum of the MOS1 and MOS2 data. The lower curve
  is off-set by a factor of 3 for clarity.  Also shown are the best-fit truncated
  spheroid plus beta models.}
  \label{fig:a1758S_josh_sb}
\end{figure}

To determine the density jump across the two edges, we extracted
surface brightness profiles within $90^{\circ}$ pie slices 
(see Fig.~\ref{fig:a1758S_regions}).
The background-subtracted and exposure-corrected surface brightness profiles
are shown in Fig.~\ref{fig:a1758S_josh_sb}. We then fit the profile across
the northern edge with a truncated spheroid plus beta model and obtained
a best-fit radius for the spheroidal component of 325~kpc,
$\beta=0.54$, and a core radius of 170~kpc.  
Fitting the same model to the
surface brightness profile across the southern edge we obtained
a best-fit radius for the spheroidal component of 200~kpc,
$\beta=0.62$, and a core radius of 350~kpc.  The best fit $\beta$ values are typical of 
5~keV clusters.  The deprojected density jumps
across the two edges are $\rho_2/\rho_1=2.1 \pm 0.3$ and $2.3 \pm 0.3$ 
(1 $\sigma$ errors) for the 
northern and southern edges, respectively.
Using the temperatures in Table 4, the pressure jumps are 
$P_2/P_1 = 1.2 \pm 0.3$ and $P_2/P_1 = 1.0 \pm 0.3$ across the northern and 
southern edges, respectively.  This shows that the pressures 
within and just beyond the truncated shperiodal components are comparable
and these features are consistent with being cold fronts.  
Given the upper limits on the pressure jumps,
the motion of these systems relative to one another other must be less than
about 1,400~km~s$^{-1}$.

As mentioned above, the present state of the merger in A1758S is 
very similar to that found in A2142 (Markevitch et al.\ 2000). 
Contrary to A1758N, the X-ray morphology of A1758S is much
more symmetric suggesting that A1758S is undergoing a
nearly head-on merger.  The morphology of A1758S is quite similar to 
that seen in the simulation of head-on merger simulations by 
Ricker \& Sarazin (2001). 
At least as seen in projection, the merger in A1758S appears
to be near the point of closest approach.

\section{Summary}

The optically rich cluster Abell 1758 is located in a dynamically active 
region that contains two hot and X-ray luminous clusters (A1758N and A1758S) 
with recessional velocities within 2,100~km~s$^{-1}$ and within 
a projected separation of 2~Mpc.
These two systems are most likely gravitationally bound, but 
there is no evidence for any interaction between the two gaseous atmospheres 
in the X-ray data. 
The X-ray morphologies of A1758N and A1758S are very complex indicating that 
both systems are undergoing major mergers. 
A1758N is in the later stages of a large impact parameter merger between 
two 7~keV clusters, while A1758S is in the earlier stages of a merger between 
two 5~keV clusters.  A1758N contains two low entropy cores that were not 
strongly shocked during the merger.  The robustness of the two cores in A1758N
indicates that the core of the dark matter distribution in the two clusters
is smaller than approximately 100 kpc. Each of these cores has a preceding cold 
front (separating the low entropy cores and shock heated gas) and a wake of low 
entropy gas.  The surface brightness profiles of the gas 
preceding the two cold fronts in A1758N are well represented by broken power-laws.  The 
slope of the outer power-law is consistent with the slope typically
found in the outskirts of rich clusters. 
The gas temperature preceding the NW subcluster in A1758N increases with 
increasing distance from the cold front.
A different trend would be observed if the break in the surface brightness
was produced by a shock.  Based on the X-ray morphology 
of A1758N, we propose that both systems have undergone significant deflections
during their merger.  Thus, these systems would have
propagated through a broad range of ambient gas pressures producing
a broad range of shock strengths.  The lower temperature gas located 
directly in front of the NW subcluster may have been shocked as the two 
cores passed close to one another while the gas pressure was high.  As
the two systems receded from one another into lower pressure gas,
the shock strength would increase producing higher post-shocked temperatures.
The chaotic entropy distribution of the gas in A1758N will drive a significant 
redistribution of the energy deposited by the merger shock via convection.

A1758S is undergoing a major merger between two 5 keV clusters.  The X-ray 
morphology of A1758S is very similar to that seen in A2142 (Markevitch et al.\ 2000).
The cores of the two subclusters are nearly coincident on the plane of the sky.
The X-ray morphology of A1758S is more symmetric than the 
morphology in A1758N, suggesting that the merger in A1758S is
more nearly head-on.  Both subclusters in A1758S are preceded by 
cold fronts, which in turn are surrounding by hotter gas that was
probably shock heated during the merger.  The upper limits on 
the pressure jump across the two cold fronts indicate that the relative 
velocity between the 
two subclusters must be less than 1,400~km~s$^{-1}$.  
The two cores in A1758S have radii of 200 kpc and 350 kpc, compared
to 80 and 115 kpc in A1758N, indicating that these cores have
not experienced as much ram pressure stripping.  This is 
consistent with A1758S being in the earlier stages of a merger,
especially if the merger in A1758S is nearly head-on.  There is 
also no evidence in the XMM-Newton data for low entropy wakes in 
A1758S, which probably indicates that all of the gas that has been 
stripped from the two cores was shocked while it was being stripped.   
This is also reasonable since ram pressure stripping should be the most
violent during the early stages of a merger while the outer, lower pressure
gas is being stripped.  

Abell 1758 is a remarkable system 
with 4 subclusters hotter than 5 keV within a region of 2~Mpc. 
Using the simulations of Randall, Sarazin, \& Ricker (2002) as a guide,
the final virialized temperature of A1758 should be about 12 keV.
We are thus witnessing the formation of a cluster which will rank among the 
hottest and most massive clusters found in the local universe.

\acknowledgments
The authors would like to thank P.E.J. Nulsen and M. Markevitch for some
very helpful discussions. This work was supported by NASA grants NAG5-12933
and AR4-5016X.

\end{document}